\documentclass[final,authoryear,5p]{elsarticle}
\usepackage{graphicx}
\usepackage[varg]{txfonts}
\usepackage{graphicx}
\usepackage{natbib}
\usepackage{lipsum}
\usepackage{amssymb}
\usepackage[%
a4paper=true,%
breaklinks=true,%
colorlinks=true,%
urlcolor=blue,
citecolor=blue,
linkcolor=blue
pdfauthor={Pereira et al.},%
pdftitle={Template for manuscripts in Advances in Space Research}%
]{hyperref}
\newcommand*\aap{A\&A }

\newcommand*\apj{ApJ }
\newcommand*\apjl{ApJ }

\newcommand*\araa{ARA\&A }

\newcommand*\jqsrt{J.~Quant.~Spectr.~Rad.~Transf.}

\newcommand*\pasj{PASJ }

\newcommand*\procspie{Proc.~SPIE}

\newcommand*\solphys{Sol.~Phys. }

\newcommand*\mdash{---}
\newcommand*\ndash{--}
\newcommand*\amp{\&}

\DeclareRobustCommand{\ion}[2]{\textup{#1\,\textsc{\lowercase{#2}}}}
\def\ha{H$\alpha$}

\def\CaII{\ion{Ca}{ii}}
\def\CII{\ion{C}{ii}}
\def\CaH{\CaII\, H}

\def\MgII{\ion{Mg}{ii}}

\def\hk{{h\&k}}
\def\MgIIhk{\MgII\, \hk}
\def\kms {{\mathrm{km}\,\mathrm{s}^{-1}}}

\journal{Advances in Space Research}

\begin{document}

\begin{frontmatter}

\title{The dynamic chromosphere: pushing the boundaries of observations and models}

\author{Tiago M. D. Pereira\corref{cor}}
\address{Rosseland Centre for Solar Physics, University of Oslo, P.O. Box 1029 Blindern, NO-0315 Oslo, Norway\\
Institute of Theoretical Astrophysics, University of Oslo, P.O. Box 1029 Blindern, NO-0315 Oslo, Norway}
\cortext[cor]{\emph{E-mail address:} \texttt{tiago.pereira@astro.uio.no}}

\begin{abstract}

The interface between the bright solar surface and the million-degree corona continues to hold the key to many unsolved problems in solar physics. Advances in instrumentation now allow us to observe the dynamic structures of the solar chromosphere down to less than $0.\!^{\prime \prime}1$ with cadences of just a few seconds and in multiple polarisation states. Such observational progress has been matched by the ever-increasing sophistication of numerical models, which have become necessary to interpret the complex observations. With an emphasis on the quiet Sun, I will review recent progress in the observation and modelling of the chromosphere. Models have come a long way from 1D static atmospheres, but their predictions still fail to reproduce several key observed features. Nevertheless, they have given us invaluable insight into the physical processes that energise the atmosphere. With more physics being added to models, the gap between predictions and observations is narrowing. With the next generation of solar observatories just around the corner, the big question is: will they close the gap?

\end{abstract}


\end{frontmatter}

\parindent=0.5 cm

\section{Introduction}

The solar chromosphere is the dynamic interface between the bright solar surface and the million-degree corona. It is an interesting object of study by itself, but also because it retains the key to understand the corona and heliosphere. Most of the photosphere's magnetoconvective energy gets deposited in the chromosphere, which regulates how this energy is converted to heat and radiation and how much energy and mass reach the solar corona.
 Studies of the chromosphere go back a very long time \citep{secchi1877} and it is not the purpose of this contribution to present a comprehensive review or give an historical perspective. Instead, our aim is to report on recent studies of the chromosphere, with a focus on the quiet Sun.

In the last decade, significant resources have been poured into understanding the chromosphere. Advances have come from pushing the boundaries of observations and, increasingly, from pushing the boundaries of numerical simulations, which have become an essential tool to understand observations. This contribution is structured around these two aspects.

\section{Pushing the boundaries of observations}

\subsection{What is the chromosphere?}

Between the photosphere and the corona lie many pressure scale heights. It is in the chromosphere that plasma ceases to be in local thermodynamic equilibrium and ionisation equilibrium. Chromospheric dynamics are governed by complex processes that are not yet fully understood. But before we concern ourselves with those complicated topics, let us take a step back and ask the most basic question: what is the chromosphere? %

The chromosphere is the layer where plasma $\beta$ hovers around unity, where matter transitions from being dominated by plasma pressure to being dominated by magnetic pressure. This has profound consequences on the structuring and morphology of the atmosphere. We can define the chromosphere phenomenologically in terms of its morphological features \citep[e.g. mottles, fibrils, straws, spicules, see][]{Rutten:2006}, or in physical terms from stratified atmosphere models as the region from the temperature minimum up to the onset of the transition region. But neither definition gives us any clues regarding what exactly is chromospheric plasma or how it is organised. The difficulty in defining the chromosphere arises partly because it is difficult to observe. Only a very small fraction of the solar flux comes from the chromosphere, either in cores of very strong spectral lines or in thermal UV radiation that is out of reach of ground-based telescopes. It is dynamic in timescales of only a few seconds, and its very fine spatial scales are not yet fully resolved even with today's most powerful telescopes.
To make matters worse, \cite{Judge:2010} predict that most chromospheric plasma exists in a `bulk, stratified chromosphere' that is not seen in chromospheric broadband images, and \cite{Rutten:2017} argue that the fibrils seen in \ha, a prime diagnostic of the chromosphere, do not trace actual plasma, but past trajectories (or contrails) of heating events. In short, we don't seem to see most of the chromosphere, and much of what we `see' may not actually be there. No wonder the chromosphere is confusing.

\begin{figure*}
\begin{center}
\includegraphics[width=\textwidth]{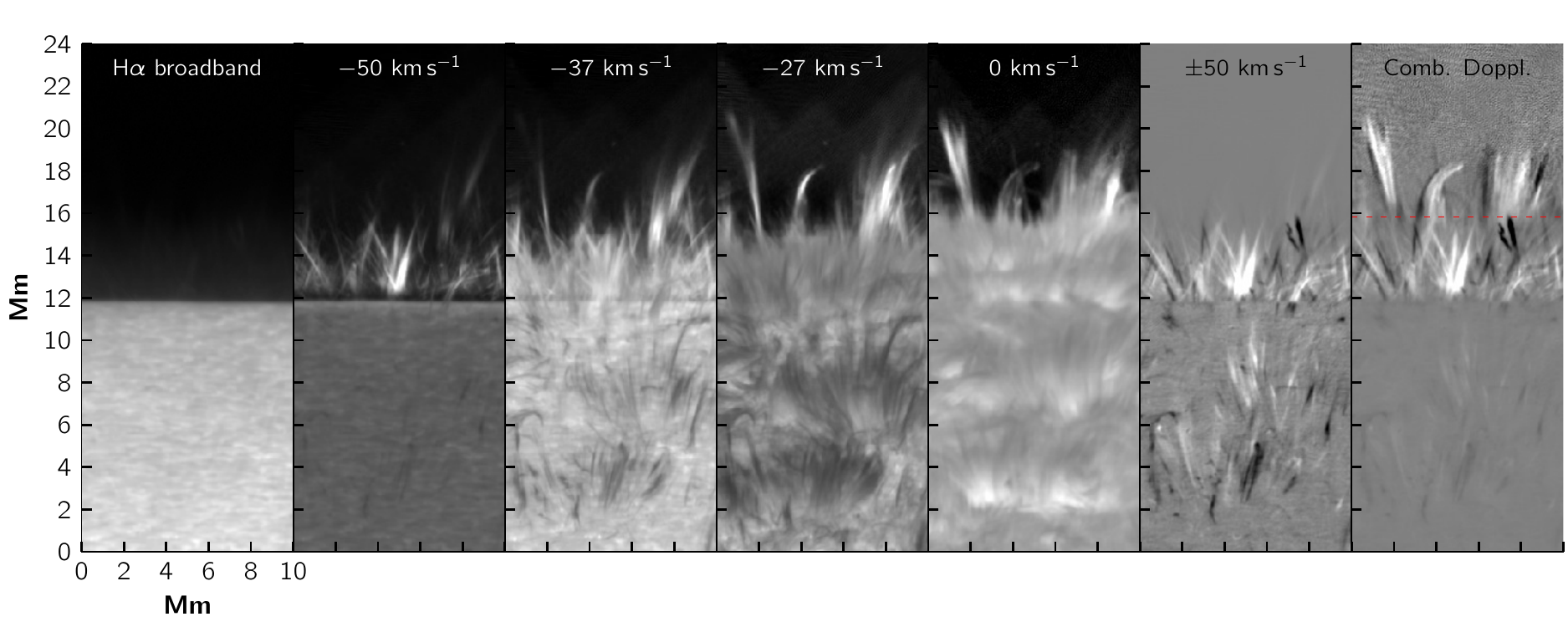}
\end{center}
\caption{Spicules and the chromosphere observed by the SST/CRISP in H$\alpha$. From left to right, panels show a broadband filter (photosphere), followed by four narrowband filters at different positions of the blue wing ($-50$ to $-27\;\kms$) and line core, a $\pm50\;\kms$ Dopplergram and a combined Dopplergram that show the full extent of the off-limb \ha\ spicules. Adapted from \cite{Pereira:2016}.\label{fig:halpha_spic}}
\end{figure*}

What is observed as the quiet Sun chromosphere, at least since the advent of \emph{Hinode} \citep{Kosugi:2007} and its Solar Optical Telescope \citep[SOT, ][]{Tsuneta:2008}, is a sea of fibrils or spicules (here we use the term for both disk and off-limb objects), thin elongated features that cover nearly the entire surface of the Sun. Spicules may appear sparse or dense depending on the width of the filter used, or if the filter samples the wings or core of chromospheric lines \citep[see][]{Pereira:2016}, and are rooted in the magnetic network.
In Figure~\ref{fig:halpha_spic} we show examples of spicules on disk and off-limb as observed in \ha\ by the Swedish 1-m Solar Telescope \citep[SST,][]{Scharmer:2003} with the CRISP interferometer \citep{Scharmer:2008}. Observed in \CaH\ filtergrams by \emph{Hinode}/SOT, spicules appear more common and dynamic than previously thought, with a subset (`type II') reaching velocities in excess of $100~\kms$ \citep{DePontieu:2007, Pereira:2012spic}. Such extreme properties seem inconsistent with earlier measurements of spicule properties \citep[the `classical' spicules of][and references therein]{Beckers:1972}, but such discrepancies can be explained by the lower resolution of earlier instruments \citep{Pereira:2013spiclett}. The nature and origin of spicules has been fiercely debated since the findings with \emph{Hinode}. Type II spicules seem to fade from the \CaH\ images at the end of their lives, prompting suggestions of violent heating that may heat the chromosphere and corona \citep{DePontieu:2009, DePontieu:2011}. \cite{Judge:2011} suggest that the spicular chromosphere is in fact a sort of optical illusion that arises from corrugated, two-dimensional sheet-like structures related to magnetic tangential discontinuities.
\cite{Judge:2012a} and \cite{Lipartito:2014} argue that the sudden appearance of spicules on disk is evidence of the `sheet scenario', but \cite{Pereira:2016} and \cite{Shetye:2016} find that such appearances and disappearances at narrow filtergrams can be naturally explained by the spicular transverse motions and Doppler shifts.

Understanding what the chromosphere is made of, and how it mediates the transfer of energy and mass between the photosphere and corona, continues to be a central question. In the last few years, powerful new tools have been deployed to answer this question; these include not only advances in spatial and temporal resolution, but a widespread use of multi-thermal diagnostics.

\begin{figure}
\begin{center}
\includegraphics[width=0.48\textwidth]{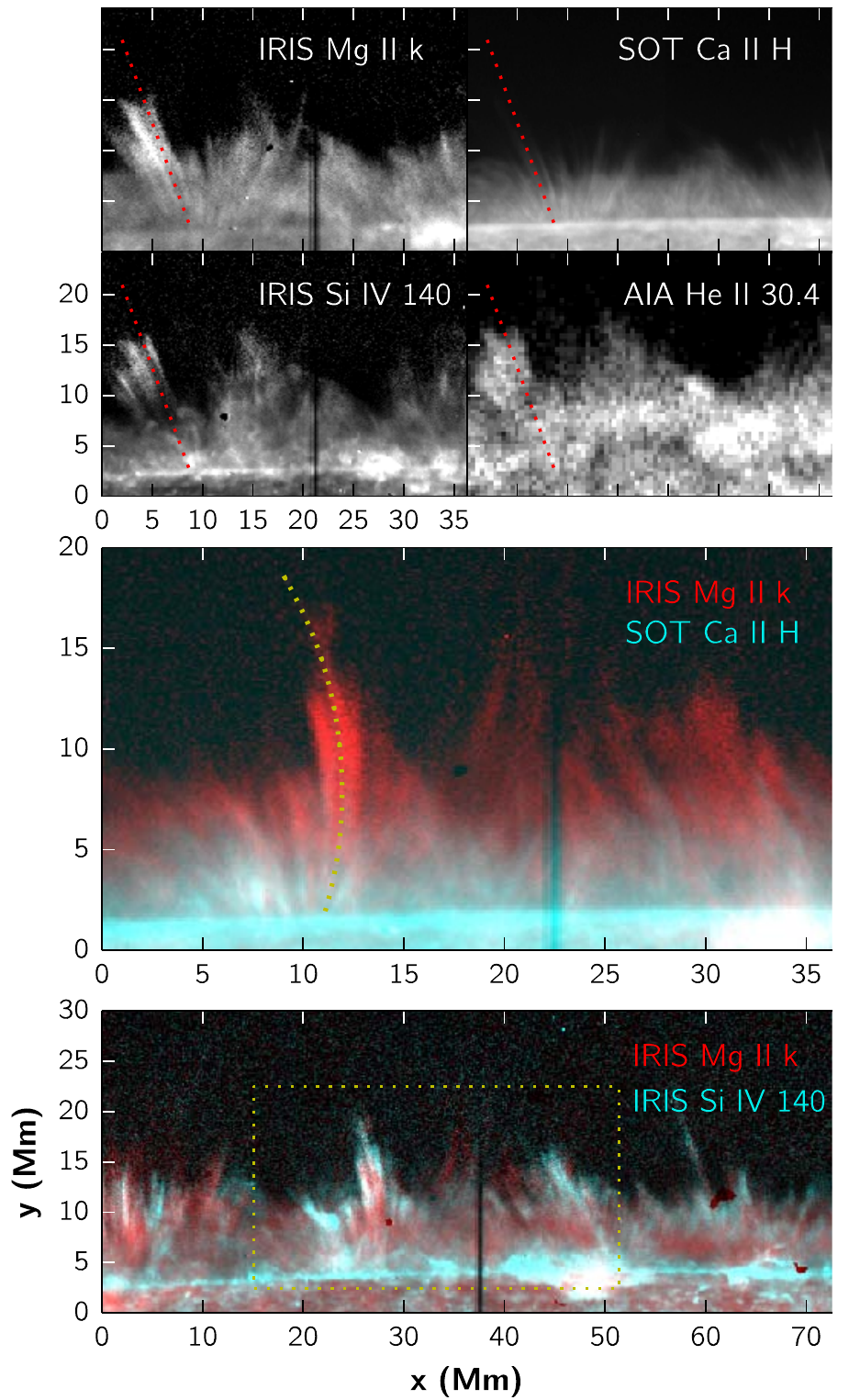}
\end{center}
\caption{A multi-thermal view of spicules. \emph{Top panel:} a quiet Sun region observed by IRIS slit-jaw filters 2796 (dominated by \MgII, formed at $\approx10,000$~K), 1400 (dominated by \ion{Si}{iv}, formed at $\approx80,000$~K), SOT \CaH\ (formed at $\approx10,000$~K), and AIA 30.4~nm filtergrams (dominated by \ion{He}{ii}, formed at $\approx100,000$~K). \emph{Middle:} superposition of IRIS 2796 (red) with SOT \CaH\ (cyan). \emph{Bottom:} superposition of IRIS 2796 (red) with 1400 (cyan) slit-jaws, where yellow dotted box represents the field of view of the middle panel. From \cite{Pereira:2014spic}.\label{fig:spic_panel}}
\end{figure}

\subsection{Multi-thermal diagnostics}

The chromosphere does not exist in isolation, and it is therefore crucial to study how chromospheric events respond to events in the photosphere and how they evolve in higher, hotter layers. This is a challenging endeavour because light from these different layers comes from a wide range of wavelengths that no single instrument can cover.
The IRIS mission \citep{IRIS-paper} was launched in 2013 with the goal of understanding how the chromosphere is coupled to the photosphere below and transition region above. Observing UV spectral lines formed over many scale heights and temperatures, it has provided some of the most detailed views of the transition region to date and a unique window into the chromosphere. It has also helped to bridge the traditional observations of the photosphere (from the ground) with those of the corona (from space), because it provides images and spectra that can be more easily matched or aligned with photospheric and coronal images.

The advantages of multi-thermal diagnostics from IRIS are evident in studies of spicules. \cite{Pereira:2014spic} combine IRIS, \emph{Hinode}, and AIA \citep{Lemen:2012} observations of spicules to trace their thermal evolution and solve the mystery of disappearing \CaH\ spicules. They find that spicules observed by IRIS keep evolving after they fade from \CaH\ images, confirming the view that spicules can be violently heated out of the \CaH\ passband. Not only is the dimming or fading of spicules from \CaH\ associated with brightenings in transition region images, there is also evidence that some spicules have a brighter top, suggesting a heating front \citep{Skogsrud:2015}. The transition region counterparts of spicules are also characterised by \cite{Tian:2014c} and \cite{Narang:2016}, who find very large apparent speeds.
\cite{vdVoort:2015} combine IRIS with the SST to find spectral signatures of heating in spicules throughout the chromosphere and transition region.

\cite{Vissers:2015} also illustrate the power of multi-thermal diagnostics by combining IRIS and the SST in studies of Ellerman bombs (EBs). The authors trace the evolution of EBs from the chromosphere to transition region and find that IRIS observations support the view of EBs as bi-directional reconnection jets. \cite{Tian:2016} use IRIS and the Chinese New Vacuum Solar Telescope to study the connection between EBs and UV bursts, a similar but hotter class of events originally discovered with IRIS by \cite{Peter:2014}.

In some cases the absence of a multi-thermal signal may be more telling than its presence. Take the example of the so-called Unresolved Fine Structure \citep[UFS,][]{Feldman:1983}, that \cite{Hansteen:2014} argue is in fact observed by IRIS in the form of small, low-lying transition region loops. Combining IRIS and the SST, \cite{Pereira:2018} find little evidence of a chromospheric heating, but find spectral signatures consistent with reconnection at transition region temperatures.

As many of these studies illustrate, combining different instruments to achieve a multi-thermal view is essential to understand the chromosphere. With the complex images and spectra we have today, multi-thermal studies are likely to become more common. Still, much remains poorly known. Density diagnostics in the quiet Sun are very difficult with optically-thick lines in the chromosphere and low signal levels in optically-thin transition region lines. Estimations of mass transfer, if possible, are limited to order of magnitude estimates. The chromospheric magnetic field, the prime mediator of energy transfer, is difficult to observe because the polarisation of light is mixed in with complex radiative transfer effects and the signal level is low with the current generation of telescopes. To interpret the increasingly complex observations it has become necessary to rely on simulations, which have also been steadily increasing in sophistication.

\section{Pushing the boundaries of models}

\subsection{Observe the model}

Inferring physical parameters from stellar spectra is a challenging problem. The Sun is no exception, even with high-quality observations that are available today. Early work on modelling atmospheres of late-type stars has relied on an approximate treatment of convection via the mixing length theory, which has several free parameters such as micro- and macroturbulence that can be adjusted to make the synthetic spectra fit the observations. To model the solar atmosphere astronomers have an extra trick, which is the use of observations at different viewing angles to build semi-empirical models of the atmosphere \citep{Holweger:1974}. In any case, all these earlier approaches have the ultimate goal of producing models of the atmosphere that can reproduce the observed spectra. Their goal is to \emph{model the observations}. About two decades ago, 3D hydrodynamic simulations of solar convection have reached a level of realism \citep{SteinNordlund1998} that they could be used to reliably predict the shapes and strengths of spectra across the solar granulation \citep{Asplund2000}, and provide robust measurements of the solar chemical composition \citep{Asplund:2009}. One advantage of this approach is that the simulations are built from first principles and have no parameters that can be freely adjusted (fudge factors) to make the synthetic spectra better fit the observations. Improvements to 3D simulations have lead to model atmospheres that better reproduce the solar spectrum \citep{Pereira:2013aa}, even compared with 1D semi-empirical models that were built to fit the solar spectrum (at least for the solar photosphere). This forward-modelling approach of using synthetic spectra from simulations and then compare them with observations to gain insight into the solar atmosphere has resulted in a paradigm shift: its focus is to \emph{observe the model}.

The last decade has been prolific in forward-modelling studies of the solar chromosphere \citep[e.g.][]{Leenaarts:2009aa, Leenaarts:2010, Leenaarts:2012halpha,Leenaarts:Mg1}. Because the magnetic field has a stronger role in the chromosphere, it is necessary to introduce a realistic field configuration in the 3D simulations, which by necessity need to be magneto-hydrodynamic (MHD). This brings in a free parameter that was previously absent. In addition, lines formed in the chromosphere are strongly affected by departures from local thermodynamical equilibrium (LTE), and the strongest lines suffer from partial redistribution (PRD). These additional complications, make the radiative transfer calculations burdensome, as we describe in \ref{sec:rt}.

\subsection{Magneto-hydrodynamic simulations and beyond}
\begin{figure*}
 \begin{center}
 \includegraphics[width=\textwidth]{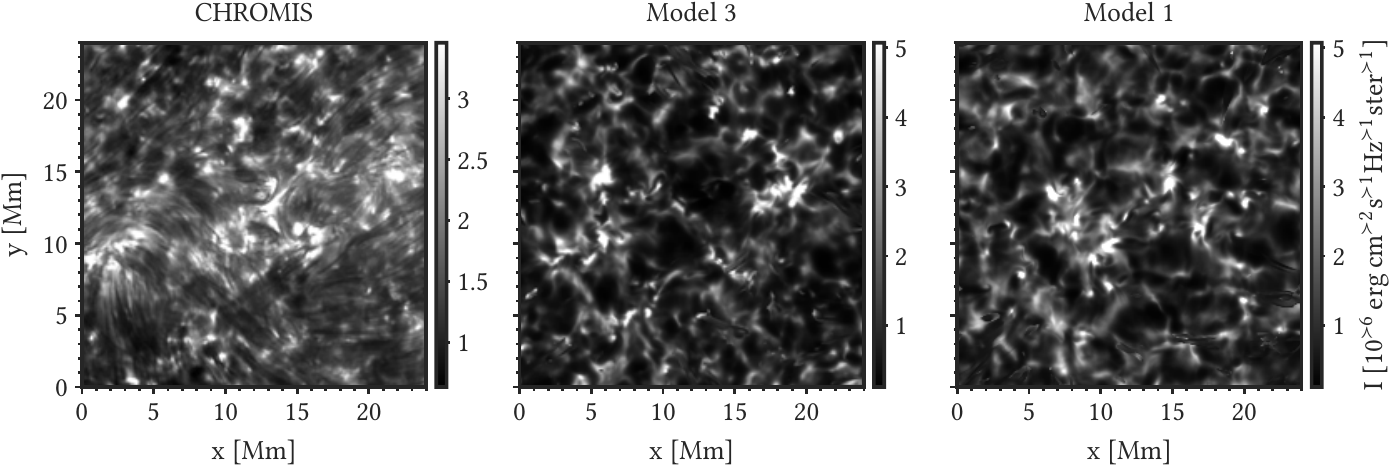}
 \end{center}
 \caption{Comparison between observed and simulated \ion{Ca}{ii}~K images at solar disk-centre. \emph{Left:} observations from SST/CHROMIS, with a spatial resolution of about $0.\!^{\prime \prime}1$ ($\approx 73$~km). \emph{Middle and right:} synthetic images from 3D non-LTE calculations from radiative MHD simulations with horizontal grid sizes of 31~km (Model 3, middle) and 95~km (Model 1, right). Image courtesy of J. Bj{\o}rgen, based on \cite{Bjorgen:2018aa}.\label{fig:chromis_sim}}
 \end{figure*}

Predicted spectra from 3D simulations of the chromosphere have not yet reached the degree of realism of their photospheric counterparts. While many predicted spectra can be similar to observations, on average they tend to be narrower and weaker, as first found for the \CaII\ 854.2~nm line \citep{Leenaarts:2009aa} and later also for the IRIS \MgII\ \citep{Leenaarts:Mg2, Pereira:Mg3} and \CII\ lines \citep{Rathore:2015b}. The reasons for this discrepancy are still unclear. In addition to spectral line strengths, chromospheric fibrils from simulations \citep[see][for examples in \ha]{Leenaarts:2012halpha} are not as abundant or as finely structured as in observations \citep{van-Noort:2006aa}, and type II spicules, one of the most conspicuous chromospheric features, are notably absent or very rare. In Figure~\ref{fig:chromis_sim}, we compare simulated and observed chromospheric images in the \CaII~K line, from \cite{Bjorgen:2018aa}. It is clear is that something is missing in the simulated chromosphere. Limits in numerical resolution may well be to blame here, as processes taking place at scales smaller than simulation cells may be responsible for additional chromospheric heating and turbulence. However, there is reason to believe that important physical effects are missing from the simulations; several new developments have, encouragingly, been narrowing the gap between observations and simulations.

\begin{figure}
\begin{center}
\includegraphics[width=0.43\textwidth]{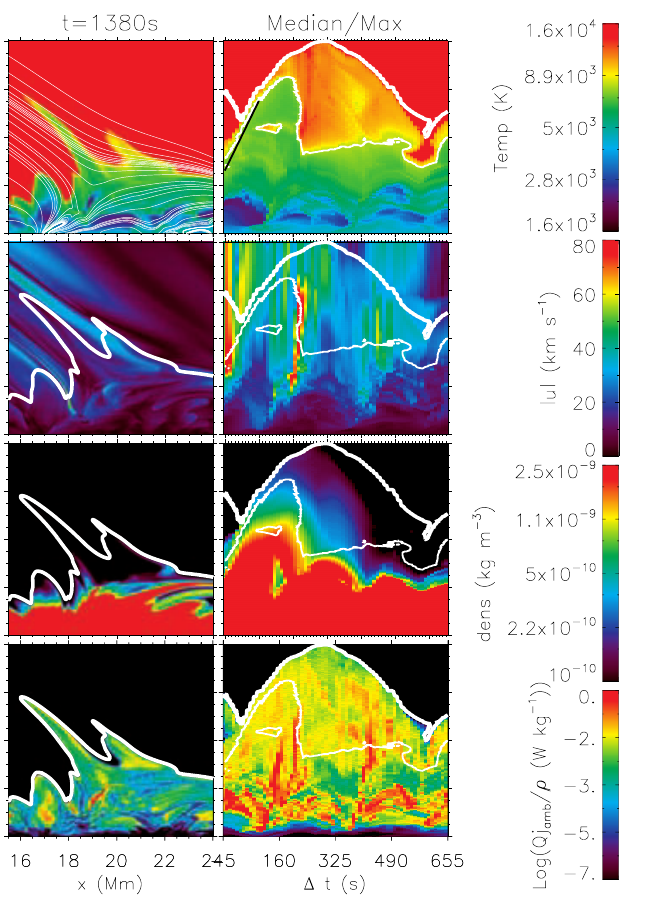}
\end{center}
\caption{Tomography of a simulated spicule. Internal quantities from a 2.5D MHD simulation showing a spicule-like jet event. From top to bottom, panels show temperature, absolute velocity, density, and median Joule heating from ambipolar diffusion. \emph{Left panels:} a side view of the simulation in the final stages of the spicule's life. \emph{Right panels:} space-time diagram of the quantities plotted along the spicule. Adapted from \cite{Martinez-Sykora:2017ab}.\label{fig:simspic}}
\end{figure}

The chromosphere is partially ionised, therefore neutrals and charged particles do not always move together. A single-fluid MHD description, used by most simulations, may be inadequate to describe the chromosphere. One option is to solve the MHD equations for several separate fluids (e.g. ions, electrons, neutrals), but this is incurs a large computational cost. Assuming that the plasma is only weakly ionised, a simpler approach is to add additional terms to the induction equation, such as the Hall term and ambipolar diffusion, and keep the single-fluid description \citep[see][and references therein]{Martinez-Sykora:2012aa}.
Using a similar prescription and stratified atmosphere models, \cite{Khomenko:2012aa} found that ambipolar diffusion can heat the chromosphere in timescales of $10-100$~s. \cite{Leake:2006aa} and \cite{Arber:2016a}, using 2.5D and 3D simulations, found that ambipolar diffusion increases the magnetic flux emergence rates. \cite{Martinez-Sykora:2012aa} introduced for the first time the Hall term and ambipolar diffusion in 2.5D dynamic simulations spanning the photosphere, chromosphere, and corona.
The authors find that ambipolar diffusion can be the dominant form of resistivity in the chromosphere, and in a follow-up work \cite{Martinez-Sykora:2017aa} conclude that `ambipolar diffusion increases the temperature in the chromosphere, [and] concentrates electrical currents, leading to more violent jets and reconnection processes.' This turns out to be quite significant, as it suggests that ion-neutral effects not only increase the temperature in the chromosphere, but lead to the creation of more violent jets, similar to type II spicules, that have been missing from self-consistent simulations. Indeed, \cite{Martinez-Sykora:2017ab} study jets from the 2.5D simulations and their predicted spectra and find a remarkable agreement with observations of type II spicules. In Figure~\ref{fig:simspic} we show different properties of simulated spicules from \cite{Martinez-Sykora:2017ab}. This was the first time that type II spicules have been reliably and systematically produced in MHD simulations, and it highlights the importance of ambipolar diffusion in amplifying concentrations of magnetic tension that later release powerful jets.

The latest dynamic MHD simulations with ambipolar diffusion still predict spectra lines that are weaker and narrower than observed, but the difference is narrowing. Future avenues include not only higher spatial resolution, but running simulations in 3D and also do explicit multi-fluid MHD calculations. These may very well be the missing pieces that the simulated chromospheres lack.

Outside first-principles simulations, idealised MHD simulations can still be of use to interpret observations of the chromosphere. For example, \cite{Antolin:2015aa} and \cite{Antolin:2018aa} study the effect of transverse MHD waves, in particular resonant absorption and Kelvin-Helmholtz instabilities, on idealised 3D MHD simulations where a flux tube is inserted, to simulate a prominence and a spicule, respectively. In both cases, the authors find that the instabilities extract energy from the resonant layer and turbulently dissipate it though vortices in the edges of the flux tube. In the case of spicules, \cite{Antolin:2018aa} conclude that transverse wave induced Kelvin– Helmholtz rolls can lead to the fine, multi-threaded, structure of spicules that is observed in chromospheric filters such as those in IRIS and \emph{Hinode}.

\subsection{Optically-thick radiative transfer\label{sec:rt}}

In the forward modelling paradigm, it is essential to do an `apples to apples' comparison between simulations and observations. This requires the calculation of synthetic spectral lines, the predicted spectra from simulations. Even though many simulations have detailed radiative cooling and heating, their internal prescription is often too approximate to synthesise individual spectral lines, let alone include effects such as non-local thermodynamic equilibrium (non-LTE) and partial redistribution (PRD), which affect many chromospheric lines. Detailed, optically-thick radiative transfer calculations must therefore be performed on the output of the simulations.

There is much to be gained by comparing the synthetic and observed spectra. We can assess the realism of the simulations, but much more importantly, we can identify similar spectral properties and learn what physical properties lead to observed signatures, because from simulations we have the full thermodynamic state of the atmosphere. This approach has proven fruitful for the IRIS mission, which observed chromospheric spectra in unprecedented detail and required guidance to interpret its complex observations. A systematic study using Bifrost simulations \citep{Gudiksen:2011, Carlsson:2016} to understand the formation of several IRIS lines has been published as a series of papers, covering the \MgIIhk\ lines \citep{Leenaarts:Mg1, Leenaarts:Mg2, Pereira:Mg3}, the \MgII\ triplet lines \citep{Pereira:2015}, the \CII\ 133.5~nm lines \citep{Rathore:2015a, Rathore:2015b, Rathore:2015c}, the \ion{O}{i} 135.56~nm line \citep{Lin:2015}, and the \ion{C}{i} 135.58~nm line \citep{Lin:2017}.
\cite{Bjorgen:2018aa} use similar models to study the formation of the \CaII~H and K lines and to help interpret the observations of CHROMIS, a new imaging spectrometer at the SST.
On a similar vein, 3D simulations were used to study the polarisation of chromospheric lines such as Ly$\alpha$ \citep{Stepan:2015aa, Schmit:2017} and \CaII\ 854.2~nm \citep{Stepan:2016aa}. These studies provide unique insight into the formation mechanism of different spectral lines, are invaluable for observers, and are a model for successfully understanding future solar observations.

So far, we have taken for granted the calculation of predicted chromospheric spectra from simulations. However, these calculations come at a big cost. Detailed, 3D non-LTE calculations are very computationally intensive and often calculations for a single atomic species take longer to run than the 3D MHD simulations themselves. Treating lines in PRD can take an order of magnitude more effort, and the fine angular grids necessary for detailed polarisation calculations can also add up one order of magnitude.
Several non-LTE codes exist today \citep[see][for a review of the methods]{Carlsson:2008}. Multi3D \citep{Leenaarts:2009a}, initially based on the MULTI code of \cite{Carlsson:1986}, has been used in various problems, from the solar abundances to chromospheric diagnostics and has been recently extended to include PRD \citep{Sukhorukov:2017} and a fast-converging multigrid scheme \citep{Bjorgen:2017}.
MUGA \citep[see][and references therein]{Fabiani-Bendicho:1997} solves the non-LTE problem using a multigrid scheme with Gauss-Seidel iterations, and has been used in several problems. The RH code \citep{Uitenbroek:2001} also works on a variety of geometries (1D, 2D, 3D, spherical) and was one of the first to allow the calculation of PRD lines. A recent modification, RH 1.5D \citep{RH15D}, only works in 1D but can calculate spectra from a 3D simulation as if each column was independent (1.5D approximation), and is publicly available. PORTA \citep{Stepan:2013} also solves the 3D non-LTE problem, but with the added advantage of being able to account for Zeeman, Hanle, and scattering polarisation. Each code with their particular strengths, they all share a common weakness: slow convergence in 3D.

The transition from 1D to 3D in non-LTE radiative transfer methods has encountered many obstacles; despite recent efforts, advances in methods are not progressing as quickly as necessary. The dynamic nature of 3D MHD atmospheres means rays have to cross regions with large inhomogeneities and gradients, introducing numerical difficulties. Convergence acceleration schemes that worked in 1D rarely do so in 3D. Difficulties or non-convergence on a single grid cell in a 3D box can stop the whole 3D calculation. More importantly, increases in the spatial resolution of simulations mean that more iterations are needed for changes to propagate, and the number of iterations required to achieve convergence increases dramatically \citep{Olson:1986}. Therefore, many 3D non-LTE studies have used either low spatial resolution simulations, or used a reduced number of cells from a larger simulation (e.g. every second grid point). Even so, 3D problems for simple atoms (5 or 6 levels) can still take several hundreds of iterations to converge, in some cases almost 1000 iterations \citep{Sukhorukov:2017}.
A quick fix is to use the 1.5D approximation, but it breaks down for radiation in the core of chromospheric lines, where the photon mean free path greatly exceeds the grid sizes \citep{Leenaarts:Mg2}. A better approach is the use of multigrid methods. \cite{Bjorgen:2017} use different resolution grids to achieve a convergence rate up 6 times faster. \cite{Steiner:2016aa} developed improved methods for polarised radiative transfer in discontinuous media, which is a promising development for dynamic 3D simulations. Still, much more effort and resources must be devoted to 3D non-LTE methods. Current methods are so computationally intensive that they are out of reach for most in the community. Despite many codes existing, the vast majority are closed-source and this leads to duplication of efforts as groups are forced to roll their own implementation. If we are to understand the complex observations of the next generation of solar observatories, the calculation of synthetic spectra is a critical piece. It desperately needs more attention.

\section{Summary}

The last few years have been prolific in studies of the solar chromosphere. Observations have increasingly focused on combining multiple spectral lines and different instruments to achieve a multi-thermal view of the chromosphere. Nevertheless, the observation of chromospheric magnetic fields remains an elusive goal. Difficulties in estimating the magnetic field and density hinder progress in our understanding of the chromosphere.

Numerical experiments and MHD simulations have evolved to become an essential tool to guide the interpretation of observations. Forward-modelling studies using detailed simulations let us understand the physical processes that lead to various observed spectral signatures. They offer a unique way to probe the magneto-thermodynamical structure of the chromosphere. Simulations fail to predict several properties of the chromosphere, but they are steadily improving. The single-fluid MHD scenario is being stretched to the limit in the partially ionised chromosphere. Simulations including ion-neutral effects show that changes in the generation and dissipation of electrical currents can lead to significant changes in chromospheric heating. Encouragingly, such simulations seem to naturally develop spicules, ubiquitously observed but previously missing in first-principle simulations. These are an important stepping stone before 3D multi-fluid simulations come of age.

As simulations evolve, so do radiative transfer calculations necessary to obtain the predicted spectra. Three-dimensional, non-LTE calculations still require a very large computational effort. Current methods struggle to cope with the highest resolution simulations, in particular when complex effects such as PRD or scattering polarisation are included. More effort is needed to efficiently solve these challenging problems, and new developments using multigrid methods and solvers optimised for discontinuous media are encouraging.

With the new generation of solar observatories just around the corner, it is imperative that we make sense of the complicated chromospheric spectra. This will require a concerted forward modelling effort aimed at reducing (or closing) the gap between the simulated and the observed chromosphere.

\vspace{0.5cm}

\noindent {\small \emph{Acknowledgements.} The author gratefully acknowledges support from the Research Council of Norway through its Centres of Excellence scheme, project number 262622.}

\vspace{0.5cm}

\bibliographystyle{elsarticle-harv}

\end{document}